\documentclass[twocolumn]{aastex62}
\usepackage{bm}

\newcommand{\lfir}{L_{\rm FIR}}
\newcommand{\lbol}{L_{\rm bol}}
\newcommand{\lsun}{L_{\odot}}
\newcommand{\ledd}{\lambda_{\rm Edd}}
\newcommand{\mbh}{M_{\rm BH}}
\newcommand{\msun}{M_{\odot}}

\received{April 3, 2019}
\revised{May 16, 2019}
\accepted{May 29, 2019}

\shorttitle{High SFRs of Low $\ledd$ Quasars at $z\gtrsim6$}
\shortauthors{Kim \& Im}

\begin{document}

\title{High Star Formation Rates of Low Eddington Ratio Quasars at $z\gtrsim6$}

\correspondingauthor{Myungshin Im}
\email{yjkim@astro.snu.ac.kr, mim@astro.snu.ac.kr}

\author[0000-0003-1647-3286]{Yongjung Kim}
\affiliation{Center for the Exploration of the Origin of the Universe (CEOU), Building 45, Seoul National University, 1 Gwanak-ro, Gwanak-gu, Seoul 08826, Republic of Korea}
\affiliation{Astronomy Program, FPRD, Department of Physics \& Astronomy, Seoul National University, 1 Gwanak-ro, Gwanak-gu, Seoul 08826, Republic of Korea}

\author[0000-0002-8537-6714]{Myungshin Im}
\affiliation{Center for the Exploration of the Origin of the Universe (CEOU), Building 45, Seoul National University, 1 Gwanak-ro, Gwanak-gu, Seoul 08826, Republic of Korea}
\affiliation{Astronomy Program, FPRD, Department of Physics \& Astronomy, Seoul National University, 1 Gwanak-ro, Gwanak-gu, Seoul 08826, Republic of Korea}



\begin{abstract}

Recent simulation studies suggest that the supermassive black hole (SMBH) growth in the early universe may precede the prolonged intense star formation within its host galaxy, instead of quasars appearing after the obscured dusty star formation phase. 
If so, high-redshift quasars with low Eddington ratios ($\ledd$) would be found in actively star-forming hosts with a star formation rate (SFR) of $>100~\msun$ yr$^{-1}$.
We present the sub-mm observations of IMS J2204+0112, 
a faint quasar with a quasar bolometric luminosity of $\lbol=4.2\times10^{12}~\lsun$ and a low $\ledd$ of only 0.1 at $z\sim6$,
carried out with the Atacama Large Millimeter/submillimeter Array (ALMA).
From its sub-mm fluxes, we measure the rest-frame far-infrared (FIR) luminosity of $\lfir=(3$--$4)\times10^{12}~\lsun$.
Interestingly, the derived host galaxy's SFR is $\sim500$--$700~\msun$ yr$^{-1}$, an order of magnitude higher than those of the $\lbol$-matched $z\gtrsim6$ quasars with high $\ledd$.
Similar FIR excesses are also found for five $z\gtrsim6$ low-$\ledd$ quasars ($\ledd<0.2$) in the literature.
We show that the overall SFR, $\mbh$, and $\ledd$ distributions of these and other sub-mm-detected quasars at $z\gtrsim6$ can be explained with the evolutionary track of high-redshift quasars in a simulation study where low $\ledd$ and high SFR quasars are expected at the end of the SMBH growth.
This suggests that the nuclear activities of the low $\ledd$, high $\lfir$ quasars are on the brink of being turned off, while their host galaxies continue to form the bulk of their stars at SFR $>100~\msun$ yr$^{-1}$.

\end{abstract}



\section{Introduction} \label{sec:introduction}

High-redshift quasars have continued to shed light on our understanding of the early universe.
To date, quasars are identified even when the universe was much less than 1 Gyr old,
with the currently known highest redshift quasar ULAS J1342+0928 at $z=7.54$ \citep{Banados18}
and hundreds of quasars discovered in the epoch of reionization from the optical/near-infrared (NIR) surveys \citep{Fan00,Fan06,Goto06,Jiang09,Jiang16,Willott10b,Mortlock11,
Venemans13,Venemans15a,Venemans15b,Banados14,Banados16,Banados18,Kashikawa15,Kim15,Wu15,
Matsuoka16,Matsuoka18,Matsuoka19,Wang16a,Wang17,Wang18a,Wang18b,Mazzucchelli17,Yang18}.
Mass estimates of supermassive black holes (SMBHs) residing at centers of these high-redshift quasars
suggest that there are SMBHs as massive as $10^{8}$--$10^{10}~\msun$ just hundreds of millions of years after the Big Bang
\citep{Kurk07,Kurk09,Jiang09,Willott10a,DeRosa11,Mortlock11,Jun15,Wu15,Mazzucchelli17,Banados18,Kim18,Onoue19,Shen19}.
Their accretion rates are found to reach the Eddington limit for most of $z\gtrsim6$ bright quasars, meaning that they are in a rapidly growing phase \citep{Willott10a,DeRosa11,DeRosa14,Trakhtenbrot14,Jun15}.
However, as quasar survey limits go fainter, recent studies have revealed previously hidden population of quasars with low Eddington ratios ($\ledd$), raising a possibility that the $\ledd$ distribution of $z\gtrsim6$ quasars is not so much different from that of lower redshift quasars \citep{Mazzucchelli17,Kim18,Onoue19,Shen19}.

Not only the central black holes (BHs) but also the dust components of their host galaxies have also been examined, which are observable at from infrared (IR) to sub-mm wavelengths.
The fraction of quasars without hot dust emission (dust temperature of $T_{d}<1,500$ K) is found to increase with redshift \citep{Jiang10,Jun13}, indicating the expeditious SMBH growth prior to the star formation at high redshift.
In the case of cool dust emission ($T_{d}<60$ K), the recent sub-mm observations of high-redshift quasars
have revealed that their rest-frame Far-infrared (FIR) luminosities ($\lfir$) are found to span a large range \citep{Petric03,Wang08,Wang10,Wang13,Wang16b,Venemans12,Venemans16,Venemans17c,Venemans18,Omont13,Willott13,Willott15,Willott17,Banados15,Mazzucchelli17,Decarli18,Izumi18,Izumi19},
inferring that their star-formation rates (SFRs) are between 10 and 2000 $\msun$ yr$^{-1}$.
These high SFR values imply that high-redshift quasar host galaxies are also growing vigorously, like ultra-luminous infrared galaxies (ULIRGs) at low redshift.

In order to grow to a SMBH weighing over $10^{9}~\msun$ hosted by a ULIRG-like galaxy in a short time of sub-Gyr, the BH accretion rate must be kept high until $z\sim6$, despite of negative feedbacks from starbursts.
Recent simulations describe this process in detail (e.g., \citealt{Li07,Sijacki09,Pezzulli16,Smidt18}).
For example, \cite{Smidt18} find that the $10^{5}~\msun$ seed BH grows with cold gas inflow and mergers to $10^{10}~\msun$ at $\ledd\lesssim1$, in succession with starburst activities in the host.
At $\mbh \sim 10^{9}~\msun$, the BH growth slows down due to feedback mechanisms, but the starburst activities are maintained a few Myrs more at several hundred $\msun$ yr$^{-1}$ due to the efficient cooling of the gas with newly synthsized metals and continued cold gas inflow.
At this later stage of the extended star-forming period, one expects to see quasars to have high $\mbh$, high SFR, but low $\ledd$.
Overall, the expected evolutionary track of this simulated quasar is to start from low $\mbh$, low SFR, high $\ledd$ to become a high $\mbh$, high SFR, and low $\ledd$ quasar.
This is somewhat of a contrast to the popular evolutionary scenario of Active Galactic Nuclei (AGN) where galaxies grow in obscured starburst via mergers, SMBHs grow rapidly at $\ledd \sim1$ and blow away the obscuring gas, and become type 1 quasars that we find in low redshift (e.g., \citealt{DiMatteo05,Springel05,Hopkins08,Hickox09,Lapi14}).

For the high-redshift quasar evolutionary picture to be true, one must find low $\ledd$ quasars with high SFR and $\mbh$.
However, it is only recently that different groups started to report the discovery of low $\ledd$ quasars at $z\gtrsim6$.
IMS J2204+0112 is a quasar at $z=5.926$ with a low bolometric luminosity of $\lbol=4.24\times10^{12}~\lsun$ \citep{Kim15,Kim18} identified from the Infrared Medium-deep Survey (IMS; M. Im et al, in preparation).
IMS is a $J$-band NIR imaging survey of the extragalactic field of which the image depth reaches $J_{\rm AB}\sim23$ mag over 120 deg$^{2}$ areas.
This quasar has $\mbh=1.23\times10^{9}~\msun$, and $\ledd=0.11$, making it one of the lowest $\ledd$ quasars among $z\gtrsim6$ quasars identified so far.
We have obtained sub-mm data of IMS J2204+0112, using the Atacama Large Millimeter/submillimeter Array (ALMA), in order to measure SFR of its host galaxy. Together with 5 other sub-mm-detected low $\ledd$ quasars in the literature, we examine if their FIR property is consistent with the evolutionary scenarios of high-redshift quasars that have been put forward lately. 

This paper is organized as follows.
We describe the ALMA observation of IMS J2204+0112 in Section \ref{sec:observations}, and present the sub-mm continuum maps of IMS J2204+0112 and its $\lfir$ measurements in Section \ref{sec:results}.
In Section \ref{sec:discussion}, we describe the FIR excess of IMS J2204+0112 and the evolution of such low-$\ledd$ quasars at high redshift, inferred from their observed characteristics.
Throughout this paper, we used the cosmological parameters of $\Omega_{m}=0.3$, $\Omega_{\Lambda}=0.7$, and $H_{0}=70$ km s$^{-1}$ Mpc$^{-1}$, which are supported by observations in the past decades (e.g., \citealt{Im97})

\begin{figure*}
\centering
\epsscale{1.1}
\plotone{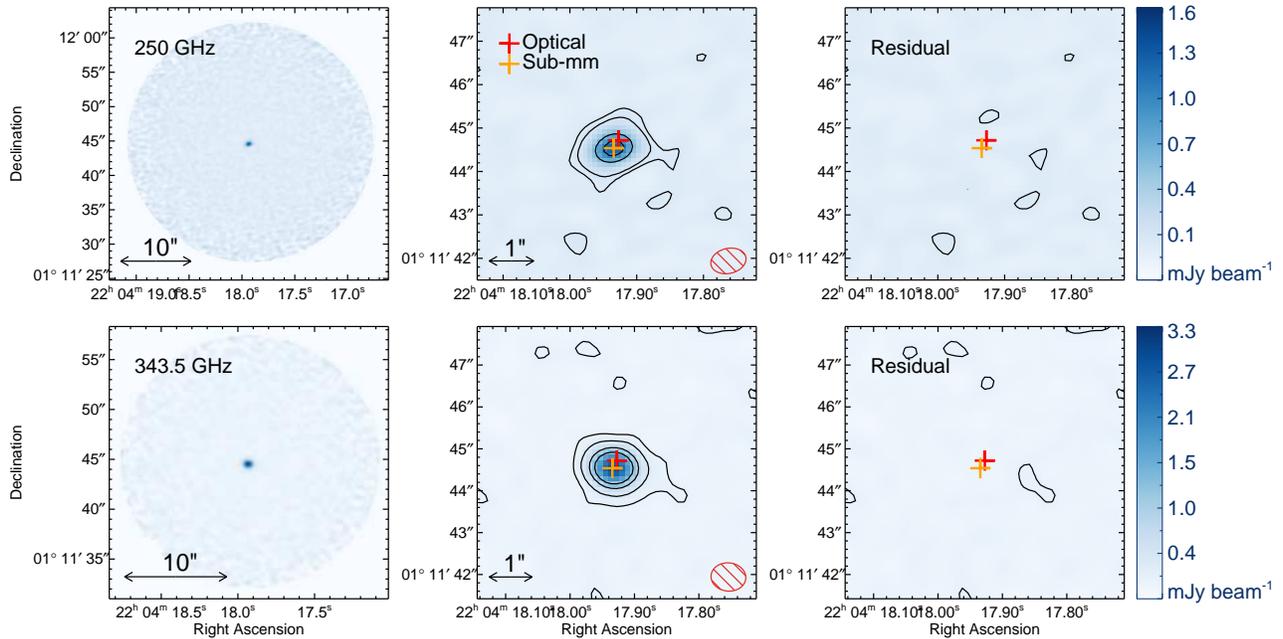}
\caption{
From left to right, ALMA integrated continuum maps covering the ALMA FOV, maps covering the central region around IMS J2204+0112, and residual maps after the 2D Gaussian model subtraction.
Top and bottom panels represent maps at 250 (band 6) and 343.5 GHz (band 7), respectively.
In both frequency maps, IMS J2204+0112 is detected as a point source without any significant neighbors.
The red and orange crosses show the positions of IMS J2204+0112 in optical (HSC-SSP $z$-band) and sub-mm (ALMA band 6), respectively,
showing the positional offset of only $\sim 0\farcs21$.
The synthetic beam sizes are given in the red ellipses in the corner.
The $1\sigma$ rms noises of the maps at 250 and 343.5 GHz are 21 and 26 $\mu$Jy, respectively, 
meanwhile the black contours indicate 2, 10, 30, and 50$\sigma$ significance levels.
\label{fig:alma}}
\end{figure*}

\section{Observations and Data} \label{sec:observations}

\subsection{ALMA} \label{sec:alma}

The ALMA observations of IMS J2204+0112 were carried out in band 6 and 7.
The band 6 data were obtained on 2016 December 13 and 2017 April 25 in the ALMA Cycle 4 project 2016.1.01311.S,
and the band 7 data were obtained on 2018 May 17 in the ALMA Cycle 5 project 2017.1.00125.S.
In both cases, 38 to 46 of the 12 m antennae were used and the baseline lengths were between 15 and 460 m,
giving an angular resolution of $0\farcs6$-$0\farcs7$.
The sources for the flux/bandpass/pointing calibration were J2148+0657 and J2253+1608, 
while J2156-0037 was observed as a phase calibrator.

Four basebands, each with a bandwidth of 1875.00 MHz and a resolution of 15.625 MHz, were used for estimating the continuum flux density integrated over a continuum bandwidth of 7.5 GHz.
The central frequencies of the bands 6 and 7 were set to 250 and 343.5 GHz, respectively.
The on-source integration times were 57.46 (band 6) and 47.88 minutes (band 7).

We used the reduced data that were provided by the ALMA Science Pipeline.
These data were processed through the standard reduction procedure of the Common Astronomy Software Application package (CASA; \citealt{McMullin07}). 
Note that the data were provided as integrated continuum maps at 250 and 343.5 GHz over the entire bandwidths and continuum maps at 4 spectral windows (basebands) with a $\sim2$ GHz bandwidth for each; 241, 243, 257, and 259 GHz for the band 6 and 336.5, 338.4, 348.5, and 350.5 GHz for the band 7.
Figure \ref{fig:alma} shows the ALMA integrated continuum maps of IMS J2204+0112.
Note that the synthesized beam sizes of the bands 6 and 7 are $0\farcs80 \times 0\farcs57$ and $0\farcs81 \times 0\farcs67$, respectively, shown as the red-hatched ellipses in the middle panels of the figure.
The rms noise values are 0.021 (band 6) and 0.026 mJy (band 7) over the 7.5 GHz bandwidth.

\subsection{Ancillary Data} \label{sec:data}

There are imaging datasets from several surveys covering IMS J2204+0112 over a wide wavelength range: the Canada-France-Hawaii Telescope Legacy Survey (CFHTLS; \citealt{Hudelot12}), IMS, the Data Release 1 of the Hyper Suprime-Cam Subaru Strategic Program (HSC-SSP DR1; \citealt{Aihara18a,Aihara18b}), the VIPERS Multi-Lambda Survey (VIPERS-MLS; \citealt{Moutard16}), the Wide-field Infrared Survey Explorer (WISE; \citealt{Wright10}), and the Faint Images of the Radio Sky at Twenty Centimeters Survey (FIRST; \citealt{Becker95}).
Among the photometric data taken at multiple epochs over the past decades,
we use the most up-to-date photometric data considering the potential variability of IMS J2204+0112 \citep{Kim18}.
For example, we used the $i$-, $z$-, and $y$-band data of HSC-SSP instead of the $i$-, $z$- and $Y$-band data of CFHTLS and IMS that were taken a few years before the HSC-SSP data.
We measured the fluxes of IMS J2204+0112 with SExtractor \citep{Bertin96} as described in \cite{Kim15,Kim19}.
Table \ref{tbl:ancillary} lists the multi-wavelength datasets and the measured flux densities.
If not detected, we used 5$\sigma$ detection limits for point sources.

\begin{deluxetable}{cccc}
\tabletypesize{\scriptsize}
\tablecaption{Flux Densities of IMS J2204+0112 from Archival Data \label{tbl:ancillary}}
\tablewidth{0pt}
\tablehead{
\colhead{Data} & \colhead{Band} & \colhead{$\lambda_{\rm obs}$ } & \colhead{$f_{\nu}$ }\\
\colhead{} & \colhead{} & \colhead{($\mu$m)} & \colhead{(mJy)} \\
\colhead{(1)} & \colhead{(2)} & \colhead{(3)} & \colhead{(4)}
}
\startdata
CFHTLS & $u'$ & 0.35 & $<1.3\times10^{-4}$ \\
CFHTLS & $g'$ & 0.48 & $<0.8\times10^{-4}$ \\
CFHTLS & $r'$ & 0.62 & $<1.6\times10^{-4}$ \\
HSC-SSP & $i$ & 0.77 & $(1.4\pm0.5)\times10^{-4}$ \\
HSC-SSP & $z$ & 0.89 & $(3.4\pm0.2)\times10^{-3}$ \\
HSC-SSP & $y$ & 0.98 & $(3.6\pm0.4)\times10^{-3}$ \\
IMS & $J$ & 1.25 & $(3.8\pm0.4)\times10^{-3}$ \\
VIPERS-MLS & $K_{s}$ & 2.15 & $(4.0\pm1.5)\times10^{-3}$ \\
WISE & $W1$ & 3.4 & $<0.068$ \\
WISE & $W2$ & 4.6 & $<0.098$ \\
WISE & $W3$ & 12 & $<0.86$ \\
WISE & $W4$ & 22 & $<5.4$ \\
FIRST & 1.4 GHz & $2.1\times10^{5}$ & $<0.95$ \\
\enddata
\tablecomments{
(1) The name of the survey from which the data was acquired.
(2) The name of the band.
(3) Observed wavelength given in units of $\mu$m.
(4) Flux density in units of mJy, except for the FIRST catalog detection limit given in units of mJy beam$^{-1}$.
}
\end{deluxetable}

\begin{figure}
\centering
\epsscale{1.20}
\plotone{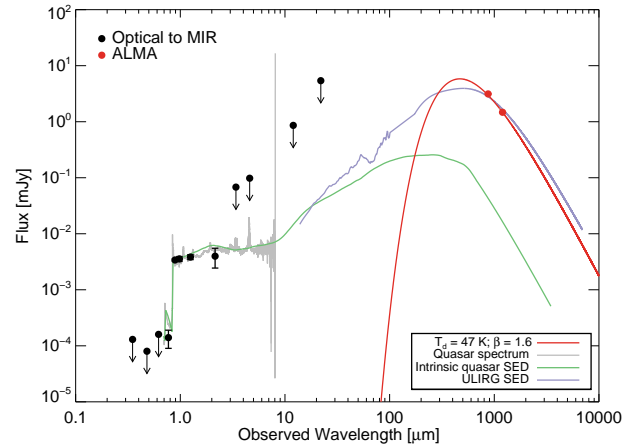}
\caption{
SED of IMS J2204+0112 in the observed frame. 
The red filled circles represent the flux densities obtained by our ALMA observation,
while the black ones are the data points from \cite{Kim15} and the archival data (see Section \ref{sec:data} and Table \ref{tbl:ancillary}).
Note that the arrows indicate the 5$\sigma$ detection limits for the undetected fluxes.
The gray, green, and purple solid lines are the SED templates of the composite quasar spectrum \citep{Selsing16},
the intrinsic SED of type 1 quasar \citep{LyuRieke17}, and the empirical SED of ULIRGs hosting AGNs at $z\sim2$ (AGN4 of \citealt{Kirkpatrick15}), respectively.
The templates are redshifted to $z=5.926$ \citep{Kim18}.
The modified blackbody model fitted for the single $f_{\rm 250GHz}$ with the fixed values of $T_{d}=47$ K and $\beta=1.6$ is shown as the red solid line (see details in Section \ref{sec:firlum}).
\label{fig:sed}}
\end{figure}

\section{Results} \label{sec:results}

\subsection{Sub-mm Continuum Maps of IMS J2204+0112} \label{sec:cont}

As shown in Figure \ref{fig:alma}, IMS J2204+0112 was clearly detected in the 250 and 343.5 GHz continuum maps obtained with ALMA (S/N $\sim60$ and 110, respectively).
There are no noteworthy objects adjacent to IMS J2204+0112, and we found no spectral features with respect to the velocity as one can expect from its redshift\footnote{
At $z=5.926$, the prominent [\ion{C}{2}] 158 $\mu$m line is located at the band gap between the band 6 and 7.
In the defined spectral windows, there could be a highly excited CO($J=$21--20) line and several H$_{2}$O lines, but they are expected to be weak and/or rare at $z\sim6$ \citep{Narayanan08,Banados15}. }. 
Using the IMFIT task of the CASA package, we fitted the source on each continuum map with a simple 2D Gaussian model, resulting in the integrated flux densities at 250 and 343.5 GHz are $f_{\rm 250GHz}=1.474\pm0.023$ and $f_{\rm 343.5GHz}=3.132\pm 0.028$ mJy, respectively.
Note that the peak flux densities are $1.289\pm0.020$ and $2.966\pm0.027$ mJy beam$^{-1}$, respectively.
These flux densities are higher than the value expected from the relation between $\lbol$ and $\lfir$ of other high-redshift quasars (equation (2) in \citealt{Venemans16}; see details in Section \ref{sec:firexcess}) by a factor of 6, although there has been a recent suggestion that there is no correlation between $\lbol$ and $\lfir$ \citep{Venemans18}.
Assuming that the FIR flux is dominated by the host galaxy, no features in the residual maps after the point source model subtraction (right panels of Figure \ref{fig:alma}) is consistent with its host galaxy being as compact as $\lesssim0\farcs7$
(or about 4 kpc in physical scale at $z\sim6$), like those of other high-redshift quasars \citep{Wang13,Willott15,Willott17,Venemans16,Venemans17a,Mazzucchelli17,Decarli18}.
The central positions of the ALMA detection are offset by only about $0\farcs2$ from the $z$-band position (see crosses in Figure \ref{fig:alma}).
These small offsets between the optical and sub-mm detections are in agreement with the previously reported uncertainties of ALMA astrometry \citep{Capak15,Willott15,Pentericci16}, disfavoring the possibility that the sub-mm flux comes from a neighboring or foreground galaxy.

Figure \ref{fig:sed} shows the spectral energy distribution (SED) of IMS J2204+0112 in the observed frame.
The black filled circles represent the flux densities of IMS J2204+0112 from \cite{Kim15,Kim18} and the values derived from the archival data (see Section \ref{sec:data}),
while the red filled circles are from our ALMA observation.
Also plotted are the composite quasar spectrum (gray line; \citealt{Selsing16}), the intrinsic SED of type 1 quasar (green line; \citealt{LyuRieke17}) and the empirical SED of ULIRGs hosting AGN at $z\sim2$ (purple line; AGN4 of \citealt{Kirkpatrick15}).
The ULIRG AGN template is consistent with the sub-mm data, which suggests that the host of IMS J2204+0112 is ULIRG-like, similar to the hosts of other high-redshift quasars \citep{Wang13,Willott13,Willott15,Willott17,Venemans16,Decarli18,Izumi18}.
Note that the templates were redshifted to the observed frame using $z=5.926$ \citep{Kim18},
including the Intergalactic Medium (IGM) attenuation effect \citep{Madau96}, and were scaled to our data points.

\subsection{FIR Luminosity and Star-formation Rate} \label{sec:firlum}

\begin{figure}
\centering
\epsscale{1.2}
\plotone{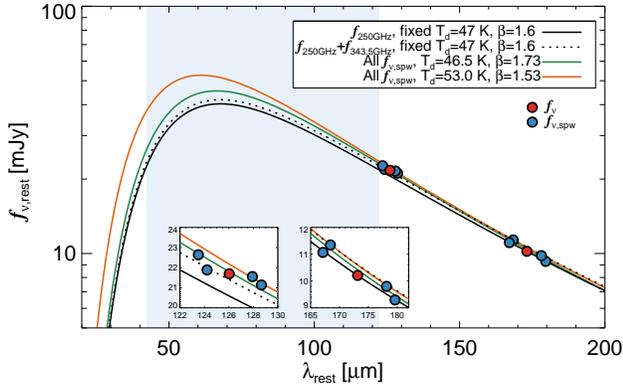}
\caption{
FIR SED of the cool dust components of IMS J2204+0112 in the rest frame. The red and blue filled circles are the $f_{\nu}$ and $f_{\nu\rm,spw}$ values, respectively.
The best-fit modified blackbody models for the $f_{\rm 250GHz}$ and $f_{\rm 250GHz}$+$f_{\rm 343.5GHz}$ with $T_{d}=47$ K and $\beta=1.6$ are shown as the black solid and dotted lines, respectively.
The green and orange lines represent the best-fit models for the $f_{\nu\rm,spw}$ values with the two sets of $T_{d}$ and $\beta$, which are from a bimodal bivariate distribution in the $T_{d}$-$\beta$ parameter space. 
The shaded region indicates the wavelength range to determine $\lfir$ (from 42.5 to 122.5 $\mu$m).
The insets are enlarged diagrams in band 6 and 7.
\label{fig:lfir}}
\end{figure}

\begin{figure}
\centering
\epsscale{0.925}
\plotone{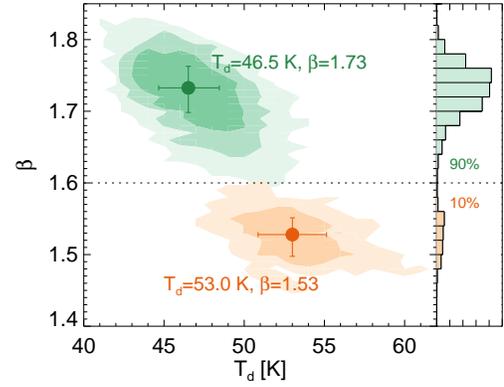}
\caption{
Posterior distribution of $T_{d}$ and $\beta$ from 10,000 trials of the Monte Carlo method described in Section \ref{sec:firlum}.
We divide the distribution by two at $\beta=1.6$ (dotted line).
The best-fit values are remarked with 1$\sigma$ errors in the panel, and also listed in Table \ref{tbl:fit}.
The contour levels represent the number of trials; 1, 10, and 50 from outer to inner.
The histograms of the divided distributions are also shown with their fractions.
\label{fig:bimod}}
\end{figure}

\cite{Dunne00} and \cite{Beelen06} suggest that the dust emission in high-redshift quasar host galaxies can be characterized by a modified blackbody model as

\begin{equation}
f_{\nu} \propto \nu^{\beta} B_{\nu}(T_{d}),
\end{equation}

\noindent where $\beta$ is the dust emissivity power-law spectral index and $B_{\nu}$ is the Planck function with a given $T_{d}$.
Following their papers, we define the $\lfir$ as the integrated luminosity over the wavelength range from 42.5 to 122.5 $\mu$m in the rest frame.
We derive $\lfir$ using several methods. 
First we estimate $\lfir$ from a single point of $f_{\rm 250GHz}$ adopting a model with fixed values of $T_{d}=47$ K and $\beta=1.6$ \citep{Beelen06}. The best-fit model using the MPFIT package \citep{Markwardt09} is shown as the black solid line in Figure \ref{fig:lfir}, resulting in $\lfir = (3.30^{+0.05}_{-0.05}) \times10^{12}~\lsun$.
Note that the uncertainty of $\lfir$ is determined by Monte Carlo method\footnote{We generated 10,000 mock sets of flux densities by adding Gaussian random noises scaled by the flux measurement uncertainties, and found a best-fit model for each set. We took a median $\lfir$ value, and the 68\% range of the inferred $\lfir$ distribution were taken as $1\sigma$ error.}.

Despite being widely used for high-redshift quasar host galaxies (e.g., \citealt{Decarli18}), the method using a single $f_{\rm 250GHz}$ with the fixed $T_{d}$ and $\beta$ values for the $\lfir$ estimation can be quite uncertain considering the wide variance of $T_{d}$ from 30 to 60 K for high-redshift quasars \citep{Beelen06,Leipski14,Venemans16,Trakhtenbrot17}.
We have continuum flux densities from as many as 8 spectral windows ($f_{\nu\rm,spw}$) in the bands 6 and 7, allowing us to trace the FIR SED of IMS J2204+0112 more accurately.
In Figure \ref{fig:lfir}, the best-fit model for two data points of $f_{\rm250GHz}$ and $f_{\rm343.5GHz}$ (red circles) with the fixed $T_{d}$ and $\beta$ is shown as the black dotted line, giving $\lfir$ of $(3.43^{+0.03}_{-0.03}) \times 10^{12}~\lsun$.
Under the same conditions, we found $\lfir$ of $(3.46^{+0.03}_{-0.03}) \times 10^{12}~\lsun$ for the eight $f_{\nu\rm,spw}$ values (blue circles).
These results are only 5\% larger than $\lfir$ from the single point of $f_{\rm 250GHz}$.

\begin{deluxetable}{cccccc}
\tabletypesize{\scriptsize}
\tablecaption{$\lfir$ and SFR of IMS J2204+0112 \label{tbl:fit}}
\tablewidth{0pt}
\tablehead{
\colhead{Band(s)} & \colhead{$T_{d}$} & \colhead{$\beta$} & \colhead{$\lfir$} & \colhead{SFR } & \colhead{Note}\\
\colhead{} & \colhead{(K)} & \colhead{} & \colhead{($10^{12}~\lsun$)} & \colhead{($\msun$ yr$^{-1}$)} & \\
\colhead{(1)} & \colhead{(2)} & \colhead{(3)} & \colhead{(4)} & \colhead{(5)} & \colhead{(6)}
}
\startdata
\multicolumn{6}{c}{Using $f_{\nu}$} \\
Band 6 & 47 & 1.6 & $\bm{3.30^{+0.05}_{-0.05}}$ & $\bm{560^{+8}_{-8}}$ & fixed $T_{d}$, $\beta$ \\
Band 6, 7 & 47 & 1.6 & $3.43^{+0.03}_{-0.03}$ & $583^{+4}_{-4}$ & fixed $T_{d}$, $\beta$ \\
\hline
\multicolumn{6}{c}{Using $f_{\nu\rm,spw}$} \\
Band 6, 7 & 47 & 1.6 & $3.46^{+0.03}_{-0.03}$ & $587^{+4}_{-4}$ & fixed $T_{d}$, $\beta$ \\
Band 6, 7 & $46.5^{+1.9}_{-1.8}$ & $1.73^{+0.03}_{-0.03}$ & $3.71^{+0.31}_{-0.33}$ & $631^{+55}_{-51}$ & $\beta>1.6$ \\
Band 6, 7 & $53.0^{+2.1}_{-2.1}$ & $1.53^{+0.02}_{-0.03}$ & $4.30^{+0.35}_{-0.37}$ & $731^{+60}_{-62}$ & $\beta<1.6$ \\
\enddata
\tablecomments{(1) the band(s) where the $f_{\nu}$ ($f_{\nu\rm,spw}$) used for fitting came from.
(2) Dust temperature in unit of K.
(3) Dust emissivity power-law spectral index.
(4) FIR luminosity determined by integrating fitted modified blackbody model from 42.5 to 122.5 $\mu$m in the rest frame.
(5) Star-formation rates estimated from FIR luminosities.
The values in bold were used for comparison with those of other quasars.
For the case with non-fixed $T_{d}$ and $\beta$, the Monte Carlo method gives a bimodal distribution of them in their parameter space, and we present the results of them in the bottom two rows (see details in Section \ref{sec:firlum}).
The reason for the small uncertainties of the cases for the fixed parameters is that the only flux measurement uncertainties are included.}
\end{deluxetable}

On the other hand, given $T_{d}$ and $\beta$ as free parameters, we found a bimodal bivariate distribution in the $T_{d}$-$\beta$ parameter space (Figure \ref{fig:bimod}).
We obtained $\lfir = (3.71^{+0.33}_{-0.31}) \times10^{12}~\lsun$ from the generated sample with $\beta>1.6$ (green contours).
In the case of $\beta<1.6$ (orange contours), we obtained $\lfir=(4.30^{+0.35}_{-0.37}) \times10^{12}~\lsun$ that is 30\% higher than the $\lfir$ from the single $f_{\rm 250GHz}$.
But the latter case accounts for only 10\% of the sample generated for the error estimation, and could be regarded as an exceptional case.

Overall, the inclusion of flux densities from more wavelengths than a single 250 GHz results in a modest increase (5--10\%, but up to 30\% in rare cases) in the $\lfir$ value.
The derived $T_{d}$ values also agree with previously reported $T_{d}$ of $z\gtrsim5$ quasars \citep{Beelen06,Leipski14,Trakhtenbrot17}.
This implies that the assumption of $T_{d}=47$ K and $\beta=1.6$ is reasonable for IMS J2204+0112 for estimating $\lfir$ to an accuracy of 5\%$-$30\%.
We listed the fitted values from the various methods in Table \ref{tbl:fit}.

Under the assumption that the FIR flux of IMS J2204+0112 mainly arises due to star formation, we estimate the SFR
following the relation of 

\begin{equation}
\frac{\rm SFR}{\msun~\rm{yr}^{-1}}\sim1.7\times 10 ^{-10}~\frac{\lfir}{\lsun},
\end{equation}

\noindent in \cite{Willott17} for the Chabrier initial mass function \citep{Carilli13}.
The SFRs estimated from the above $\lfir$ values are in the range of 560-731 $\msun$ yr$^{-1}$, and they are also listed in Table \ref{tbl:fit}.

In the following sections, we used the $\lfir$ value derived from $f_{\rm 250GHz}$ as the representative value of IMS J2204+0112, for the sake of comparison with other $z\gtrsim6$ quasars for which $\lfir$ are derived from single data points at $\sim250$ GHz.

\section{Discussion} \label{sec:discussion}

\begin{figure}
\centering
\epsscale{1.2}
\plotone{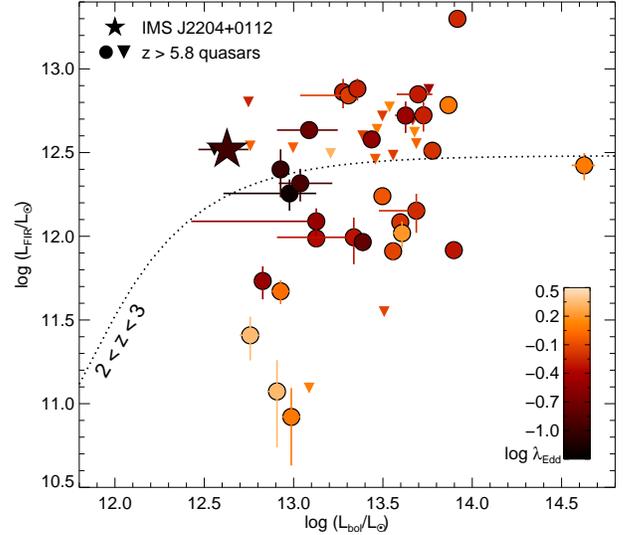}
\caption{
$\lbol$-$\lfir$ distributions of quasars.
The star symbol indicates IMS J2204+0112. 
The filled circles with error bars represent the $z>5.8$ quasars which have both UV and FIR measurements in the literature, while the upside-down triangles are the upper limits on $\lfir$ of FIR-undetected sources (see details in Section \ref{sec:firexcess}).
The colors of the symbols indicate the $\ledd$ of the quasars.
The dotted line shows the relation for quasars at $2<z<3$ \citep{Harris16}.
\label{fig:lbol-lfir}}
\end{figure}

\begin{figure*}
\centering
\epsscale{1.0}
\plotone{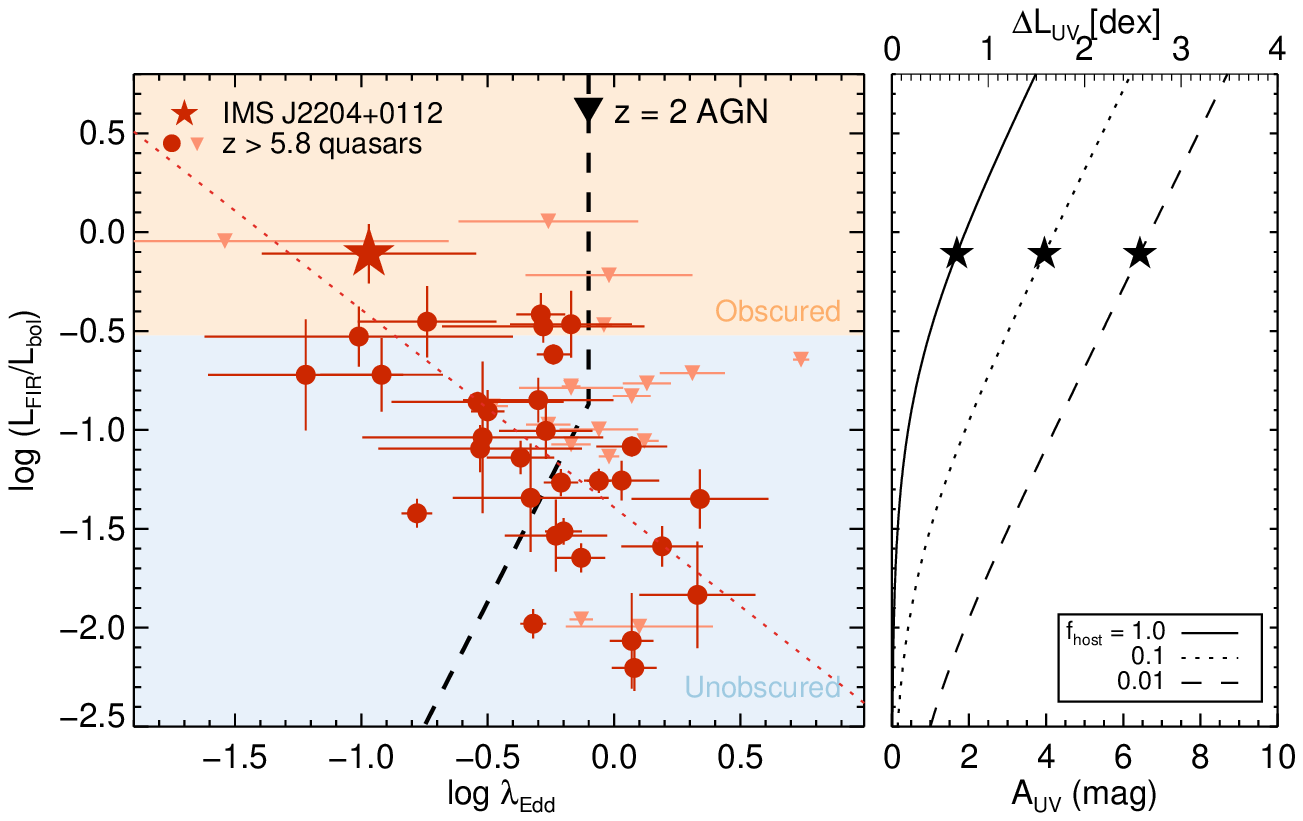}
\caption{
\emph{Left}: $\lfir/\lbol$ of high-redshift quasars with respect to $\ledd$.
The symbols are the same as in Figure \ref{fig:lbol-lfir}, while their colors are given red and orange depending on whether they were detected or not.
The red dotted line shows a negative correlation of $\lfir/\lbol\propto{\ledd}^{-1}$.
The dashed line with an arrow shows the simple evolutionary track of AGN at $z=2$ \citep{Lapi14}, while the arrow indicates the direction of evolution.
The orange/blue shaded regions for obscured/unobscured quasars are robustly divided at $\lfir/\lbol=0.3$ following \cite{Lapi14}.
\emph{Right}: $\lfir/\lbol$ as a function of UV extinction.
The solid, dotted, and dashed lines represent the cases of $f_{\rm host}=1.0$, 0.1, and 0.01, respectively.
The star symbols indicate the cases of IMS J2204+0112.
\label{fig:ledd}}
\end{figure*}

\subsection{FIR Excess of IMS J2204+0112} \label{sec:firexcess}

IMS J2204+0112 is a relatively low luminosity quasar with $\lbol=4.24\times10^{12}~\lsun$ \citep{Kim18}.
However, the observed $\lfir$ of IMS J2204+0112  is comparable to the average $\lfir$ value of other high-redshift quasars with $\lbol>10^{13}~\lsun$ ($\lfir\sim3\times10^{12}~\lsun$; \citealt{Venemans18}).
High $\lfir$ of IMS J2204+0112 is inconsistent with the previous suggestion that low-luminosity quasars are hosted by low-$\lfir$ galaxies \citep{Willott13,Willott17,Izumi18}.

For comparison, we plot in Figure \ref{fig:lbol-lfir} the $\lfir$ versus $\lbol$ values of IMS J2204+0112 (star) and other $z>5.8$ quasars (circles) that have both the rest-UV spectral properties and the rest-FIR continuum properties in the literature.
For the other quasars, The $\lbol$ values were derived from $L_{3000}$ (a luminosity at 3000 $\rm\AA$ in the rest frame) with a bolometric correction factor of 5.18 \citep{Runnoe12}.
Meanwhile, the $\lfir$ values were derived in the same manner as IMS J2204+0112 using the FIR continuum flux densities at $\sim250$ GHz in the literature (i.e. a single FIR flux density is used for each quasar).
For FIR-undetected quasars, we used 3$\sigma$ detection limits on FIR flux densities, shown as the upside-down triangles in Figure \ref{fig:lbol-lfir}.
In addition, we estimated the $\mbh$ of the quasars from their $L_{3000}$ and FWHM values of \ion{Mg}{2} emission line following \cite{Vestergaard09} under the assumption of the virial motions of \ion{Mg}{2} emitting gas, giving their $\ledd$ values as well.
The derived values are given in Table \ref{tbl:z6qsos}.

\begin{deluxetable*}{cccccccc}
\tabletypesize{\scriptsize}
\tablecaption{Derived Rest-UV and FIR Properties of $z\gtrsim6$ Quasars from the Literature \label{tbl:z6qsos}}
\tablewidth{0pt}
\tablehead{
\colhead{ID} & \colhead{$z$}  & \colhead{$\lbol$} & \colhead{$\mbh$} & \colhead{$\ledd$} & \colhead{$\lfir$} & \colhead{SFR} & \colhead{References} \\
\colhead{} & \colhead{} & \colhead{($10^{12}~\lsun$)} & \colhead{$(10^{8}~\msun)$} & \colhead{} & \colhead{($10^{12}~\lsun$)} & \colhead{($\msun$ yr$^{-1}$)} & \colhead{} \\
\colhead{(1)} & \colhead{(2)} & \colhead{(3)} & \colhead{(4)} & \colhead{(5)} & \colhead{(6)} & \colhead{(7)} & \colhead{(8)}  
}
\startdata
J0005$-$0006 & $5.844$ & $  14.4^{+  0.3}_{-  0.3}$ & $   0.8^{+  0.1}_{-  0.1}$ & $ 5.50$ & $< 3.26$ & $< 553$ & 1, 2\\
J0028$+$0457 & $5.99$ & $  27.4^{+  0.2}_{-  0.2}$ & $  28.8^{+ 26.1}_{- 17.4}$ & $ 0.29$ & $ 3.79^{+0.07}_{ -0.07}$ & $ 644\pm 12$ & 3, 4\\
J0033$-$0125 & $6.02$ & $   8.5^{+  0.1}_{-  0.1}$ & $  26.3^{+ 80.8}_{- 19.8}$ & $ 0.10$ & $ 2.51^{+0.80}_{ -0.80}$ & $ 426\pm135$ & 3, 5\\
J0050$+$3445 & $6.253$ & $  46.5^{+  4.5}_{-  5.1}$ & $  25.7^{+  4.5}_{-  4.3}$ & $ 0.55$ & $< 4.95$ & $< 840$ & 6, 7\\
J0055$+$0146 & $5.983$ & $   8.5^{+  1.0}_{-  0.9}$ & $   2.4^{+  0.8}_{-  0.7}$ & $ 1.07$ & $ 0.47^{+0.08}_{ -0.08}$ & $  79\pm 12$ & 6, 8\\
J0100$+$2802 & $6.30$ & $ 423.7^{+ 62.8}_{- 63.1}$ & $ 107.2^{+ 15.9}_{- 13.8}$ & $ 1.20$ & $ 2.65^{+0.49}_{ -0.49}$ & $ 450\pm 83$ & 9, 10\\
J0109$-$3047 & $6.763$ & $  13.4^{+  1.3}_{- 10.7}$ & $  13.5^{+  4.7}_{-  8.9}$ & $ 0.30$ & $ 1.23^{+0.24}_{ -0.24}$ & $ 208\pm 41$ & 11, 12\\
J0136$+$0226 & $6.21$ & $   5.6^{+  0.1}_{-  0.1}$ & $   3.1^{+  2.9}_{-  1.9}$ & $ 0.55$ & $< 6.34$ & $<1077$ & 3, 7\\
J0210$-$0456 & $6.438$ & $   5.7^{+  0.7}_{-  0.7}$ & $   0.8^{+  0.6}_{-  0.4}$ & $ 2.19$ & $ 0.26^{+0.07}_{ -0.07}$ & $  43\pm 12$ & 6, 13\\
J0221$-$0802 & $6.161$ & $   6.7^{+  0.6}_{-  0.7}$ & $   6.9^{+  7.5}_{-  4.6}$ & $ 0.30$ & $ 0.54^{+0.12}_{ -0.12}$ & $  91\pm 20$ & 6, 14\\
J036.5078$+$03.0498 & $6.533$ & $  53.3^{+  5.1}_{- 16.4}$ & $  30.2^{+ 11.5}_{-  9.8}$ & $ 0.54$ & $ 5.28^{+1.06}_{ -1.06}$ & $ 897\pm179$ & 11, 15\\
J0227$-$0605 & $6.21$ & $   5.7^{+  0.1}_{-  0.1}$ & $   1.8^{+  1.6}_{-  1.1}$ & $ 0.95$ & $< 3.46$ & $< 588$ & 3, 7\\
J0303$-$0019 & $6.079$ & $   9.9^{+  0.2}_{-  0.2}$ & $   3.3^{+  0.2}_{-  0.2}$ & $ 0.91$ & $< 3.38$ & $< 574$ & 1, 2\\
J0305$-$3150 & $6.61$ & $  20.3^{+  2.5}_{-  9.4}$ & $   9.1^{+  3.5}_{-  3.6}$ & $ 0.68$ & $ 6.95^{+0.21}_{ -0.21}$ & $1180\pm 35$ & 11, 12\\
J0353$+$0104 & $6.072$ & $  36.1^{+  1.7}_{-  1.6}$ & $  15.8^{+  2.8}_{-  2.7}$ & $ 0.68$ & $< 3.05$ & $< 518$ & 1, 2\\
J0836$+$0054 & $5.81$ & $  18.9^{+  1.4}_{-  1.7}$ & $  11.0^{+  2.5}_{-  2.1}$ & $ 0.51$ & $ 7.27^{+1.45}_{ -1.45}$ & $1235\pm247$ & 16, 17\\
J0841$+$2905 & $5.95$ & $  28.6^{+  0.1}_{-  0.1}$ & $  10.0^{+  3.8}_{-  3.2}$ & $ 0.87$ & $< 2.89$ & $< 490$ & 3, 5\\
J0842$+$1218 & $6.069$ & $  39.5^{+  1.9}_{-  1.8}$ & $  19.1^{+  3.3}_{-  2.8}$ & $ 0.63$ & $ 1.21^{+0.11}_{ -0.11}$ & $ 206\pm 19$ & 1, 4\\
J1030$+$0524 & $6.302$ & $  32.1^{+  0.7}_{-  0.7}$ & $  13.2^{+  1.3}_{-  1.4}$ & $ 0.74$ & $< 0.35$ & $<  60$ & 1, 4\\
J1048$+$4637 & $6.198$ & $  73.6^{+  1.7}_{-  1.7}$ & $  19.1^{+  6.6}_{-  5.6}$ & $ 1.17$ & $ 6.08^{+0.14}_{ -0.14}$ & $1033\pm 24$ & 1, 4\\
J167.6415$-$13.4960 & $6.505$ & $  12.2^{+  3.9}_{-  5.5}$ & $   3.0^{+  1.2}_{-  1.4}$ & $ 1.26$ & $< 0.12$ & $<  21$ & 11, 4\\
J1120$+$0641 & $7.087$ & $  48.6^{+  5.9}_{- 18.7}$ & $  25.1^{+  8.0}_{-  8.9}$ & $ 0.59$ & $ 1.42^{+0.37}_{ -0.37}$ & $ 241\pm 63$ & 11, 18\\
J1137$+$3549 & $6.01$ & $  57.2^{+  0.3}_{-  0.3}$ & $  52.5^{+  7.8}_{-  6.8}$ & $ 0.33$ & $< 7.54$ & $<1281$ & 3, 19\\
J1148$+$5251 & $6.407$ & $  78.9^{+  1.8}_{-  1.8}$ & $  50.1^{+  6.1}_{-  5.5}$ & $ 0.48$ & $ 0.83^{+0.10}_{ -0.10}$ & $ 140\pm 17$ & 1, 4\\
J1148$+$0702 & $6.34$ & $  36.1^{+  0.2}_{-  0.2}$ & $  14.8^{+  3.4}_{-  3.0}$ & $ 0.74$ & $ 0.81^{+0.10}_{ -0.10}$ & $ 138\pm 16$ & 3, 4\\
J1205$-$0000 & $6.73$ & $   9.5^{+  3.9}_{-  5.4}$ & $  47.9^{+ 61.8}_{- 18.4}$ & $ 0.06$ & $ 1.80^{+0.38}_{ -0.38}$ & $ 306\pm 64$ & 11, 11\\
J1207$+$0630 & $6.03$ & $  24.4^{+  0.2}_{-  0.2}$ & $  44.7^{+  6.6}_{-  5.8}$ & $ 0.17$ & $ 0.92^{+0.11}_{ -0.11}$ & $ 157\pm 18$ & 3, 4\\
J1250$+$3130 & $6.14$ & $  34.4^{+  0.2}_{-  0.2}$ & $   7.8^{+  1.8}_{-  1.6}$ & $ 1.35$ & $< 5.92$ & $<1006$ & 3, 19\\
J1306$+$0356 & $6.017$ & $  31.4^{+  0.7}_{-  0.7}$ & $  11.0^{+  1.1}_{-  1.2}$ & $ 0.87$ & $ 1.74^{+0.13}_{ -0.13}$ & $ 295\pm 21$ & 1, 4\\
J1335$+$3533 & $5.90$ & $  42.4^{+  1.0}_{-  1.0}$ & $  40.7^{+  6.0}_{-  6.1}$ & $ 0.32$ & $ 5.26^{+1.12}_{ -1.12}$ & $ 894\pm191$ & 20, 19\\
J1342$+$0928 & $7.527$ & $  40.5^{+  3.9}_{-  4.4}$ & $   7.8^{+  3.7}_{-  2.1}$ & $ 1.55$ & $ 1.04^{+0.18}_{ -0.18}$ & $ 177\pm 31$ & 21, 22\\
J1411$+$1217 & $5.903$ & $  47.5^{+  1.1}_{-  1.1}$ & $  10.7^{+  1.3}_{-  1.4}$ & $ 1.32$ & $< 4.18$ & $< 710$ & 1, 19\\
J1427$+$3312 & $6.12$ & $  29.3^{+  0.2}_{-  0.2}$ & $   7.6^{+  1.3}_{-  1.3}$ & $ 1.17$ & $< 4.35$ & $< 739$ & 3, 5\\
J1429$+$5447 & $6.12$ & $  22.8^{+  0.2}_{-  0.2}$ & $  13.2^{+ 13.7}_{-  8.9}$ & $ 0.52$ & $ 7.61^{+1.14}_{ -1.14}$ & $1292\pm194$ & 3, 7\\
J1509$-$1749 & $6.121$ & $  59.8^{+  5.8}_{-  6.5}$ & $  29.5^{+  3.6}_{-  3.2}$ & $ 0.62$ & $ 3.25^{+0.09}_{ -0.09}$ & $ 551\pm 16$ & 6, 4\\
J231.6576$-$20.8335 & $6.587$ & $  49.8^{+ 10.1}_{- 12.0}$ & $  30.9^{+  6.3}_{- 20.7}$ & $ 0.50$ & $ 7.05^{+0.10}_{ -0.10}$ & $1198\pm 17$ & 11, 4\\
J1602$+$4228 & $6.08$ & $  48.6^{+  0.2}_{-  0.2}$ & $  15.5^{+  1.5}_{-  1.4}$ & $ 0.95$ & $< 3.58$ & $< 607$ & 3, 5\\
J1623$+$3112 & $6.211$ & $  31.4^{+  0.7}_{-  0.7}$ & $  14.1^{+  1.0}_{-  1.2}$ & $ 0.68$ & $< 5.23$ & $< 888$ & 1, 19\\
J1630$+$4012 & $6.058$ & $  24.4^{+  4.9}_{-  5.0}$ & $  11.0^{+  5.3}_{-  4.0}$ & $ 0.68$ & $< 3.98$ & $< 676$ & 1, 2\\
J1641$+$3755 & $6.047$ & $  16.1^{+  1.6}_{-  1.8}$ & $   2.4^{+  0.7}_{-  0.6}$ & $ 2.04$ & $< 3.12$ & $< 530$ & 6, 7\\
J2100$-$1715 & $6.087$ & $  13.4^{+  1.3}_{-  1.5}$ & $   9.3^{+  3.0}_{-  2.4}$ & $ 0.43$ & $ 0.97^{+0.11}_{ -0.11}$ & $ 165\pm 19$ & 6, 4\\
J323.1382$+$12.2986 & $6.592$ & $  21.7^{+  1.6}_{- 13.7}$ & $  14.1^{+  4.1}_{-  7.0}$ & $ 0.47$ & $ 0.99^{+0.31}_{ -0.31}$ & $ 167\pm 52$ & 11, 11\\
J2229$+$1457 & $6.152$ & $   8.1^{+  0.8}_{-  0.9}$ & $   1.2^{+  0.7}_{-  0.5}$ & $ 2.14$ & $ 0.12^{+0.06}_{ -0.06}$ & $  20\pm 10$ & 6, 8\\
J338.2298$+$29.5089 & $6.66$ & $  10.9^{+  5.6}_{-  2.6}$ & $  27.5^{+ 12.3}_{- 10.9}$ & $ 0.12$ & $ 2.07^{+0.46}_{ -0.46}$ & $ 352\pm 77$ & 11, 11\\
J2310$+$1855 & $5.96$ & $  82.6^{+  0.2}_{-  0.2}$ & $  43.7^{+  6.5}_{-  6.5}$ & $ 0.58$ & $19.91^{+0.18}_{ -0.18}$ & $3384\pm 30$ & 3, 23\\
J2329$-$0301 & $6.417$ & $   9.7^{+  1.2}_{-  1.1}$ & $   2.5^{+  0.4}_{-  0.4}$ & $ 1.17$ & $ 0.08^{+0.04}_{ -0.04}$ & $  14\pm  6$ & 6, 14\\
J2348$-$3054 & $6.902$ & $  12.2^{+  5.4}_{-  4.1}$ & $  20.4^{+  8.4}_{-  9.7}$ & $ 0.18$ & $ 4.31^{+0.31}_{ -0.31}$ & $ 732\pm 53$ & 11, 12\\
J2356$+$0023 & $6.05$ & $   3.6^{+  0.1}_{-  0.1}$ & $  38.9^{+ 84.1}_{- 36.8}$ & $ 0.03$ & $< 3.25$ & $< 553$ & 3, 2\\
\enddata
\tablecomments{(1) ID of quasars. 
(2) Redshift from UV spectra (e.g., \ion{Mg}{2}).
(3) Bolometric luminosity.
(4) Black hole mass.
(5) Eddington ratio ($\lbol/L_{\rm Edd}$).
(6) FIR luminosity.
(7) Star-formation rate.
(8) References for rest-UV and rest-FIR, respectively:  1---\cite{DeRosa11};  2---\cite{Wang11};  3---\cite{Shen19};  4---\cite{Decarli18};  5---\cite{Wang08};  6---\cite{Willott10a};  7---\cite{Omont13};  8---\cite{Willott15};  9---\cite{Wu15};  10---\cite{Wang16b};  11---\cite{Mazzucchelli17};  12---\cite{Venemans16};  13---\cite{Willott13};  14---\cite{Willott17};  15---\cite{Banados15};  16---\cite{Kurk07};  17---\cite{Petric03};  18---\cite{Venemans12};  19---\cite{Wang07};  20---\cite{Eilers18};  21---\cite{Banados18};  22---\cite{Venemans17c};  23---\cite{Wang13}.
}
\end{deluxetable*}

It is remarkable that the $\lfir$ value of IMS J2204+0112 is an order of magnitude higher than that of its $\lbol$-matched quasar (CFHQS J0210$-$0456; \citealt{Willott10a,Willott13}), which has $\ledd\sim2$.
Likewise, the recently discovered $z\gtrsim6.5$ quasars with low $\ledd$ \citep{Mazzucchelli17,Shen19} also have higher $\lfir$ values than those of their $\lbol$-matched sample with high $\ledd$.
This trend is more prominent in Figure \ref{fig:ledd} which shows a negative correlation between $\lfir/\lbol$ and $\ledd$ of the high-redshift quasars  $(\lfir/\lbol\propto{\ledd}^{-1}; red dotted line)$, although these quasars are not a complete sample.
Note that we cannot find such a negative correlation for the Palomar-Green (type 1) Quasars \citep{Lani17,Lyu17}.
In particular, IMS J2204+0112 has the highest $\lfir/\lbol$ value of 0.8 among the sources with sub-mm detection in Figure \ref{fig:ledd}.
If it were at a low redshift, this quasar can be classified as an obscured quasar that is in the evolving stage before the optically bright type 1 quasar phase ($\lfir/\lbol>0.3$; \citealt{Hao05,Lapi14,Mancuso17}).

Since the $\lbol$ of IMS J2204+0112 is derived from its UV continuum luminosity, one may argue that the large $\lfir/\lbol$ ratio is a result of absorption/scattering of the UV flux by the dust in its host galaxy.
We examine if the dust absorption is the reason for its low luminosity and $\ledd$.
Under the assumption that its host galaxy is a starburst galaxy, we estimated the UV extinction of the host galaxy ($A_{\rm UV}$ at 0.16 $\mu$m) of IMS J2204+0112 from the ratio of the host galaxy's FIR and UV luminosities (equation (7) in \citealt{Calzetti00}):

\begin{equation}
A_{\rm UV}\simeq 2.5 \log \left[ \frac{1}{0.9} \frac{\lfir}{ f_{\rm host} L_{\rm UV} } +1 \right],
\end{equation}

\noindent where $L_{\rm UV}$ is the UV luminosity at 0.16 $\mu$m following the prescription of \cite{Runnoe12}, and $f_{\rm host}$ is the fractional contribution of the host galaxy to $L_{\rm UV}$. 
Here, we also assume that $\lfir$ is dominated by the host galaxy.

In the right panel of Figure \ref{fig:ledd}, we show the change of $A_{\rm UV}$ in terms of $\lfir/\lbol$.
The lower limit of the UV extinction of IMS J2204+0112 would be $A_{\rm UV}>1.7$ or $E(B-V)>0.4$, which is achieved when $f_{\rm host}=1$ (solid line).
Application of the $A_{\rm UV}>1.7$ correction would increase the intrinsic $\lbol$ of IMS J2204+0112 by $>0.7$ dex, which in turn gives $\lfir/\lbol<0.1$, in agreement with the $\lfir/\lbol$ values of type 1 quasars \citep{Lapi14,Lani17,Lyu17,Stanley17} and the $\lbol$-$\lfir$ relation of $z\gtrsim6$ quasars (Figure 7 and equation (2) in \citealt{Venemans16}).
However, the suggestion that IMS J2204+0112 is an obscured quasar can be rejected due to the following reasons.
First, IMS J2204+0112 has evident Ly$\alpha~\lambda1216$ and \ion{C}{4} $\lambda1549$ emission lines \citep{Kim18}.
Given such a large $E(B-V)$ value, the UV emission lines are expected to be weak or undetectable even in luminous quasars ($\lbol>3\times10^{12}~\lsun$; \citealt{Wethers18}).
Second, the spectrum of IMS J2204+0112 shows a moderate UV power-law slope of $\alpha_{\lambda}=-1.12$ \citep{Kim18}, inconsistent with the expectation for an obscured quasar.
For the large $A_{\rm UV}$ value, the intrinsic $\alpha_{\lambda}$ should be much steeper than $\alpha_{\lambda}<-3.5$ that is a rare case for quasars.
Finally, the above situations become worse if $f_{\rm host}<1$.
For example, $A_{\rm UV}$ increases to 6.4 if we assume that 1 \% of the UV photons are from its host galaxy ($f_{\rm host}=0.01$; dashed line).
In fact, the host-to-AGN UV flux ratio of quasars with $\lbol>10^{12}~\lsun$ is almost zero \citep{Shen11}, and $A_{\rm UV}$ becomes extremely high ($\gg6.4$) in such a case.

\begin{figure*}
\centering
\epsscale{1.0}
\plotone{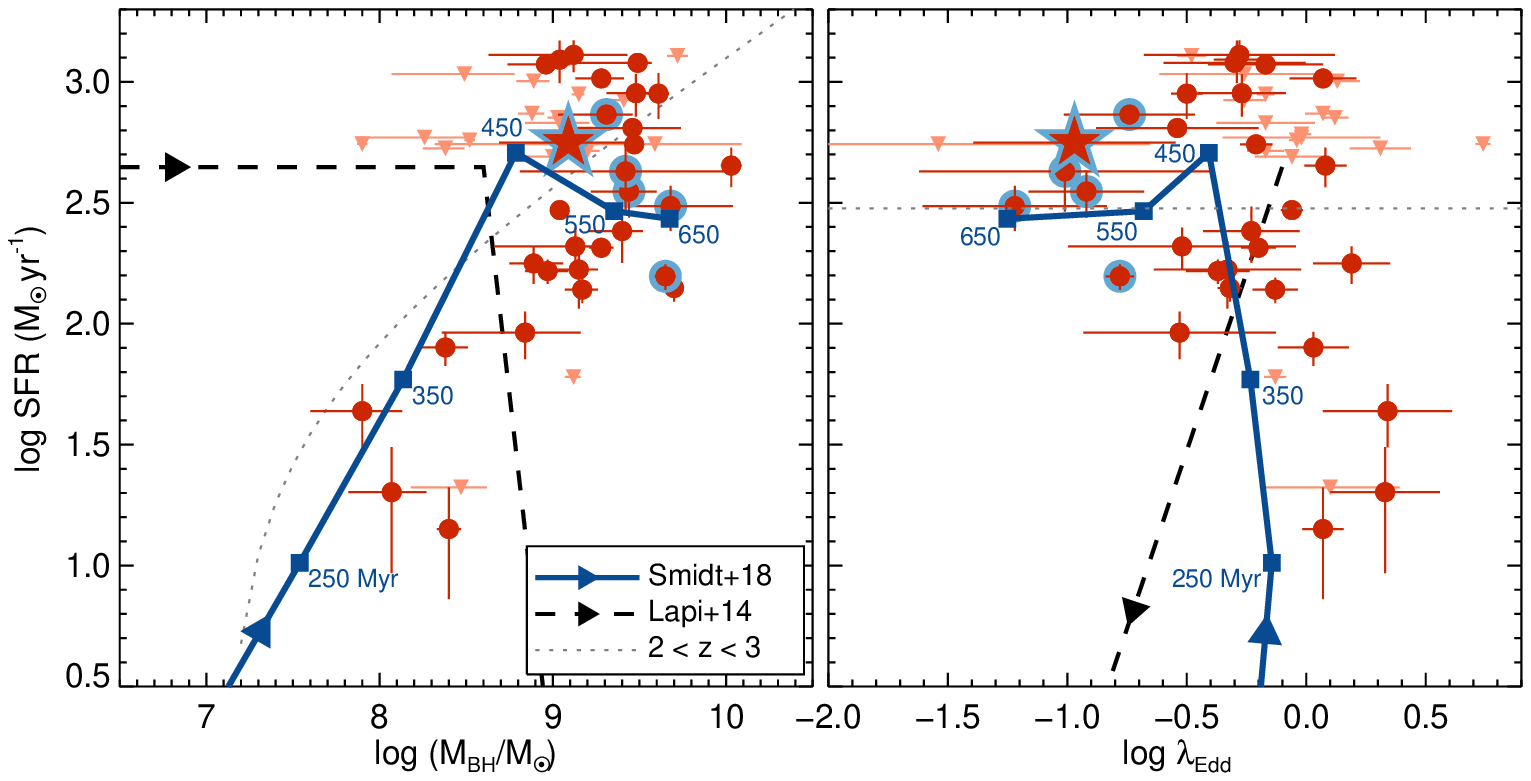}
\caption{
SFRs of high-redshift quasars along $\mbh$ (left) and $\ledd$ (right).
The symbols are same as Figure \ref{fig:ledd}, while the LEQs ($\ledd<0.2$) are highlighted with blue outlines.
The evolutionary track of high-redshift quasars by \cite{Smidt18} is shown as the navy solid lines with arrows indicating the direction of evolution, while the denoted numbers are the time since a $10^{5}~\msun$ seed BH began to grow.
The dashed lines with arrows show the simple evolutionary track of AGN at $z=2$ \citep{Lapi14}.
The dotted line in the left panel is a power-law model matched to quasars at $2<z<3$ \citep{Harris16}.
Note that in the right panel, the mean SFR value of 300 $\msun$ yr$^{-1}$ is plotted for quasars at $2<z<3$ since there was no obvious correlation between their $\ledd$ and SFRs.
\label{fig:sfr}}
\end{figure*}

One possibility is that dust is not along our line of sight, allowing us to see its central engine.
It may happen under the assumption of the spaciously distributed dust components \citep{Lyu18}, where the dust along the polar direction (or the line of sight) was blown out by strong outflows from the central BH.
For example, there are optically selected $0.5<z<4$ quasars that are also FIR detected with high $\lfir/\lbol$ values, although they occupy only a few percents of the whole sample of optically selected quasars \citep{Pitchford16}.

Like IMS J2204+0112, the spectral features of other $z>5.8$ quasar sample we used also show a little possibility of being obscured by the dust in their host galaxies. Therefore, in the following discussions, we regard the estimated $\ledd$ values of them as intrinsic ones without any UV extinction.

\subsection{SMBH Activity and  Star Formation} \label{sec:coevolution}

In the previous section, we found a negative correlation between the $\ledd$ and $\lfir/\lbol$ of high-redshift quasars.
This correlation is mainly because of the FIR excesses of low-$\ledd$ quasars ($\ledd<0.2$, hereafter referred to as LEQ), including IMS J2204+0112. 
A mere conjecture for the FIR excesses is that their relatively weak SMBH activities are not enough to efficiently quench the star formation within their host galaxies.
But such a simple picture is inadequate to explain the widely spanned $\lfir$ of high-$\ledd$ quasars.

A currently popular scenario for the co-evolution of quasars and host galaxies is that obscured star-formation occurs first (possibly triggered by galaxy merger), followed by a blowout phase, and then to type 1 quasar and finally normal galaxies after the type 1 quasar activity subsides (e.g., \citealt{DiMatteo05,Springel05,Hopkins08,Hickox09,Netzer09,Lapi14}).
According to this scenario, quasars start to be identified in the blowout phase as somewhat obscured quasars with high $\ledd$ and SFRs \citep{Hao05,Glikman07,Georgakakis09,KimD15,KimD18}.
Then, later they become low to moderate $\ledd$ quasars in low SFR hosts.
Following this, we expect LEQs at $z\gtrsim6$ to have low SFR hosts, but on contrary, they are found to be in high SFR hosts (see the blue-outlined symbols in Figure \ref{fig:sfr}), and yet its dust obscuration is minimal.

This unexpected property of LEQs can be explained as the end stage of quasar evolution in the early universe as put forward in recent simulation works.
In Figure \ref{fig:sfr}, we plot the evolutionary track of a BH in the simulation of \cite{Smidt18}, shown as the navy solid lines with arrows indicating the direction of evolution.
Note that we binned the track into 100 Myr for simplification.
This track shows the growth of a direct collapse BH ($10^{5}~\msun$) fed by cold and dense streams to an SMBH as massive as $10^{10}~\msun$ at $z\sim6$, while the star formation within its host galaxy is boosted by mergers and metal enrichments at an epoch coeval to or later than the time when the rapid BH growth occurred.
At the end phase, the accretion rate subsides to $\ledd\sim0.1$, while the SFR is maintained at a few hundreds of $\msun$ yr$^{-1}$. 
This end stage of quasars in the simulation result is consistent with the characteristics of the LEQs, suggesting that the central engines of these LEQs could be in the end game, while their host galaxies are expected to grow further.

It is also noteworthy in Figure \ref{fig:sfr} that the evolutionary track of \cite{Smidt18} is in line with the distributions of not only the LEQs but also the other high-redshift quasars on the diagrams.
In this view, the low SFR of some high $\ledd$ quasars (e.g., J0210$-$0456, J2229$+$1457, and J2329$-$0301) are because they are too young to start the intense starbursts with metal enrichments.
This suggestion of their young ages is also supported by their sizes of proximity zone, which are smaller than the sizes expected from $M_{1450}$ \citep{Eilers17}.
If this overall picture of the quasar evolution applies to the majority of $z\sim6$ quasars, we expect that there will be very few $\ledd\sim0.1$ quasars with low SFRs at $z\gtrsim6$.
Future deep sub-mm observation of more $\ledd\sim0.1$ quasars at $z\gtrsim6$ should teach us if this is the case.

Finally, we caution that the $\mbh$-SFR distribution of $z\sim6$ quasars is in line with that of $2<z<3$ quasars (dotted line; \citealt{Harris16}).
The high-$\ledd$ quasars with low SFRs can also be explained by the episodic super-Eddington accretion that suppresses the star formation in host galaxies \citep{DeGraf17}, leaving a possibility that high-redshift quasar evolution is much more diverse than the simple picture we discussed earlier.

\section{Summary} \label{sec:summary}

In this paper, we present the sub-mm observations of IMS J2204+0112, a faint $z\sim6$ quasar with $M_{1450}=-24$ mag, using ALMA. 
We also examine if the observed sub-mm property of this and other high-redshift quasars agrees with recent simulation results.
Followings are what we find in this work.

\begin{enumerate}
\item We obtained the 250 and 343.5 GHz (band 6 and 7, respectively) continuum maps of IMS J2204+0112 by ALMA, which show detections with S/N of 60 and 110, respectively. We find that IMS J2204+0112 has flux densities of $f_{\rm 250GHz}=1.5$ mJy and $f_{\rm 343.5GHz}=3.1$ mJy. 
\item Assuming the modified blackbody model for cool dust, we estimate the $\lfir$ of (3.30--4.30)$\times10^{12}~\lsun$ for IMS J2204+0112, or the SFR of 560--731 $\msun$ yr$^{-1}$. The inclusion of the band 7 data slightly increases the $\lfir$ by 10\% with $T_{d}=46.5$ K and $\beta=1.73$ (but up to 30\% in rarely extreme situations). This implies that the widely used cool-dust model for high-redshift quasars with $T_{d}=47$ K and $\beta=1.6$ using a single $f_{\rm 250GHz}$ is a suitable assumption for IMS J2204+0112. 
\item We find that the derived $\lfir$ of IMS J2204+0112 is high in comparison to that of quasars with similar $\lbol$ ($\lfir/\lbol=0.8$ versus $<0.1$). At low redshift such high $\lfir/\lbol$ quasars are mostly dust-obscured quasars. However the spectral features of IMS J2204+0112 rule out the possibility of this quasar being highly obscured. 
\item The FIR excesses are also found for other five low-$\ledd$ quasars ($\ledd<0.2$) in the literature. Combined with other quasars with higher $\ledd$ and sub-mm detection, the overall distribution of the high-redshift quasars in the $\mbh$, $\ledd$, and SFR ($\lfir$) space is consistent with simulation results of quasars in the early universe, where low $\ledd$ and high SFR quasars are expected near at the end of the SMBH growth.
\end{enumerate}

Since the number of low-$\ledd$ quasars used in the discussion is small, enlarging the sample is necessary to see the validity of our suggestion.
The recently reported low $\ledd$ quasars at $z\gtrsim5.7$ \citep{Shen19} can be good candidates for deep sub-mm observations with ALMA, allowing us to judge whether quasars with low $\ledd$ and SFRs exist or not.
Also, there are a handful number of high-redshift quasars with extremely large $\lfir/\lbol$ ratios ($>0.3$ or beyond; \citealt{Venemans18} and references therein), but without $\mbh$ and $\ledd$ measurements.
Deep NIR spectroscopy of such objects, possibly with upcoming future facilities such as Giant Magellan Telescope and/or James-Webb Space Telescope, should shed light on the general properties of high $\lfir$ quasars.

\acknowledgments

This work was supported by the National Research Foundation of Korea (NRF) grant, 
No. 2017R1A3A3001362, funded by the Korea government (MSIP).

This paper makes use of the following ALMA data: ADS/JAO.ALMA\#2016.1.01311.S and \#2017.1.00125.S. ALMA is a partnership of ESO (representing its member states), NSF (USA) and NINS (Japan), together with NRC (Canada), MOST and ASIAA (Taiwan), and KASI (Republic of Korea), in cooperation with the Republic of Chile. The Joint ALMA Observatory is operated by ESO, AUI/NRAO and NAOJ.

Based on observations obtained with MegaPrime/MegaCam, a joint project of CFHT and CEA/IRFU, and with WIRCam, a joint project of CFHT, Taiwan, Korea, Canada, France, at the Canada-France-Hawaii Telescope (CFHT) which is operated by the National Research Council (NRC) of Canada, the Institut National des Science de l'Univers of the Centre National de la Recherche Scientifique (CNRS) of France, and the University of Hawaii. 
This work is based in part on data products produced at Terapix available at the Canadian Astronomy Data Centre as part of the Canada-France-Hawaii Telescope Legacy Survey, a collaborative project of NRC and CNRS.

The United Kingdom Infrared Telescope (UKIRT) is supported by NASA and operated under an agreement among the University of Hawaii, the University of Arizona, and Lockheed Martin Advanced Technology Center; operations are enabled through the cooperation of the Joint Astronomy Centre of the Science and Technology Facilities Council of the U.K.

The Hyper Suprime-Cam (HSC) collaboration includes the astronomical communities of Japan and Taiwan, and Princeton University. The HSC instrumentation and software were developed by the National Astronomical Observatory of Japan (NAOJ), the Kavli Institute for the Physics and Mathematics of the Universe (Kavli IPMU), the University of Tokyo, the High Energy Accelerator Research Organization (KEK), the Academia Sinica Institute for Astronomy and Astrophysics in Taiwan (ASIAA), and Princeton University. Funding was contributed by the FIRST program from Japanese Cabinet Office, the Ministry of Education, Culture, Sports, Science and Technology (MEXT), the Japan Society for the Promotion of Science (JSPS), Japan Science and Technology Agency (JST), the Toray Science Foundation, NAOJ, Kavli IPMU, KEK, ASIAA, and Princeton University. 
This paper makes use of software developed for the Large Synoptic Survey Telescope. We thank the LSST Project for making their code available as free software at  http://dm.lsst.org
The Pan-STARRS1 Surveys (PS1) have been made possible through contributions of the Institute for Astronomy, the University of Hawaii, the Pan-STARRS Project Office, the Max-Planck Society and its participating institutes, the Max Planck Institute for Astronomy, Heidelberg and the Max Planck Institute for Extraterrestrial Physics, Garching, The Johns Hopkins University, Durham University, the University of Edinburgh, Queen’s University Belfast, the Harvard-Smithsonian Center for Astrophysics, the Las Cumbres Observatory Global Telescope Network Incorporated, the National Central University of Taiwan, the Space Telescope Science Institute, the National Aeronautics and Space Administration under Grant No. NNX08AR22G issued through the Planetary Science Division of the NASA Science Mission Directorate, the National Science Foundation under Grant No. AST-1238877, the University of Maryland, and Eotvos Lorand University (ELTE) and the Los Alamos National Laboratory.
Based in part on data collected at the Subaru Telescope and retrieved from the HSC data archive system, which is operated by Subaru Telescope and Astronomy Data Center at National Astronomical Observatory of Japan.

This research has made use of the NASA/ IPAC Infrared Science Archive, which is operated by the Jet Propulsion Laboratory, California Institute of Technology, under contract with the National Aeronautics and Space Administration.
This publication makes use of data products from the Wide-field Infrared Survey Explorer, which is a joint project of the University of California, Los Angeles, and the Jet Propulsion Laboratory/California Institute of Technology, funded by the National Aeronautics and Space Administration.


\vspace{5mm}
\facilities{ALMA}

\software{SExtractor \citep{Bertin96},
CASA \citep{McMullin07}
}





\appendix

\section{Non-detections with SCUBA-2} \label{sec:jcmt}

IMS J2204+0112 was also observed with Submillimetre Common-User Bolometer Array 2 (SCUBA-2) on the James Clerk Maxwell Telescope (JCMT) operated by East Asian Observatory (PID: M18AP016), on 2018 June and July (5 nights) under the dry weather conditions; $0.03\leq \tau_{\rm 225GHz} \leq 0.09$ and the average seeing of $\sim1\farcs0$.
The data were simultaneously obtained at 450 and 850 $\mu$m with the on-source integration time of 4.17 hours, and the rms sensitivities are 20.16 and 0.95 mJy beam$^{-1}$ for 450 and 850 $\mu$m data, respectively.
However, we do not identify IMS J2204+0112 in the SCUBA-2 images, due to their shallow depths.
Therefore, the SCUBA-2 data are excluded from the analysis of the result.

\end{document}